\documentclass[prb,preprint]{revtex4-1} 
\usepackage{amsmath}  % needed for \tfrac, \bmatrix, etc.
\usepackage{amsfonts} % needed for bold Greek, Fraktur, and blackboard bold
\usepackage{graphicx} % needed for figures

\begin{document}

% Be sure to use the \title, \author, \affiliation, and \abstract macros
% to format your title page.  Don't use lower-level macros to  manually
% adjust the fonts and centering.

\title{A Minimal Thermo–Fluid Model for Pressure-Driven Extraction in a Moka Pot}
% In a long title you can use \\ to force a line break at a certain location.

%When submitting the manuscript for review, do not include the author's name or institution
%\author{}
\author{Syahril Siregar}
\email{syahril.siregar@ui.ac.id} % optional
\altaffiliation{Department of Physics, Faculty of Mathematics and Natural Sciences, Universitas Indonesia} 
% See the REVTeX documentation for more examples of author and affiliation lists.

\date{\today}

\begin{abstract}
The moka pot provides a familiar example of a thermally driven flow system in which heating, vapor pressure generation, and fluid extraction are strongly coupled. We present a minimal, dimensionless dynamical model describing the evolution of temperature, pressure, and extracted volume during moka pot brewing. The model consists of a small set of coupled ordinary differential equations incorporating constant heating, heat loss, vapor pressure buildup, and pressure-driven flow through the coffee bed. The heating stage of the model is quantitatively compared with published experimental temperature–time data, allowing the characteristic thermal timescale to be fixed independently. Using the experimentally constrained temperature evolution as input, the model predicts the pressure rise and identifies the onset of extraction without additional fitting parameters. Despite its simplicity, the model exhibits several qualitatively distinct extraction regimes, including delayed onset of flow, smooth extraction, and rapid extraction driven by nonlinear feedback between temperature and pressure. These regimes are governed by a small number of dimensionless parameters with clear physical interpretation. Rather than providing detailed quantitative predictions for specific devices, the model is intended as a transparent pedagogical framework for illustrating how physicists construct, simplify, and test coupled thermo-fluid models using experimentally accessible data in an everyday physical system in an everyday physical context.
\end{abstract}
% AJP requires an abstract for all regular article submissions.
% Abstracts are optional for submissions to the "Notes and Discussions" section.

\maketitle % title page is now complete

\section{Introduction}

Household coffee brewing devices provide familiar examples of coupled thermal and fluid processes that are accessible to a broad audience yet rich in underlying physics~\cite{bossart2025science}. Among these devices, the moka pot offers a particularly instructive system in which heating, phase change, pressure buildup, and pressure-driven flow interact in a confined geometry~\cite{navarini2009experimental}. Despite its simplicity and widespread use, the physical mechanisms governing moka pot extraction are rarely discussed in a quantitative framework suitable for physics education or systematic analysis.

From a physical perspective, the operation of a moka pot involves several classical processes: heating of a liquid in a partially closed vessel, vapor pressure generation, threshold-driven flow through a porous medium, and mass and energy transport between coupled compartments~\cite{gianino2007experimental}. These elements span thermodynamics, fluid mechanics, and transport phenomena, making the moka pot a concrete and familiar system through which these concepts may be explored~\cite{king2008physics}.

Previous studies of coffee brewing have largely focused on drip filtration or espresso-style extraction, often emphasizing empirical optimization or chemical composition rather than first-principles modeling~\cite{cameron2020systematically}. While detailed computational approaches have been employed in some cases, their complexity limits pedagogical accessibility. In contrast, low-dimensional models based on ordinary differential equations can offer substantial physical insight while remaining analytically and numerically tractable, making them particularly suitable for instructional use.

The physics of moka pot operation has been analyzed in detail by King (2008), who developed a thermofluid description based on mass and energy balances in the boiler and flow through the coffee bed~\cite{king2008physics}. That analysis highlights the physical origin of extraction, emphasizing vapor pressure buildup, hydraulic resistance, and positive feedback leading to rapid or unstable flow. While physically grounded, such balance-law approaches remain closely tied to device-specific parameters and dimensional formulations.

Building on these ideas, we introduce a deliberately reduced, fully dimensionless model expressed in terms of a small number of coupled ordinary differential equations. Each dimensionless parameter admits a clear physical interpretation. Unlike prior dimensional treatments, the present formulation emphasizes a dynamical-systems perspective that separates distinct physical regimes and enables systematic exploration of extraction behavior using a minimal parameter set. The model captures qualitatively distinct regimes, including delayed onset of flow, smooth extraction, and rapid extraction driven by nonlinear feedback between temperature and pressure.

A central goal of this work is to demonstrate how such a minimal model can be confronted with experimental data in a transparent and pedagogically meaningful way. The heating-stage temperature evolution predicted by the model is quantitatively compared with published experimental measurements, allowing the characteristic thermal timescale to be fixed independently. Using the experimentally constrained temperature as input, the model then predicts the pressure rise and identifies the onset of extraction without additional fitting parameters. This separation between calibration and prediction highlights how simplified models can be tested and refined using limited experimental information.

Beyond its relevance to coffee preparation, the moka pot thus serves as a compact pedagogical example of coupled nonlinear thermo-fluid dynamics in an everyday physical system. The model is readily implementable using standard numerical tools and can be adapted for use in advanced undergraduate courses or computational laboratories.

The remainder of this paper is organized as follows. Section~II introduces the dimensionless thermo--fluid model and its governing equations. Section~III presents numerical results for the coupled temperature, pressure, and extraction dynamics. Section~IV compares the model with experimental data and discusses its limitations and possible extensions.

\section{Methods}

\subsection{Modeling Philosophy and Scope}

The moka pot extraction process involves coupled thermal, thermodynamic, and fluid transport phenomena, including heating, vapor pressure generation, pressure buildup, and pressure-driven flow through a porous coffee bed. A fully resolved description would require spatially dependent, multiphase flow equations coupled to heat transfer, which would obscure the essential mechanisms and limit pedagogical accessibility.

\begin{figure}
    \centering
    \includegraphics[width=0.6\linewidth]{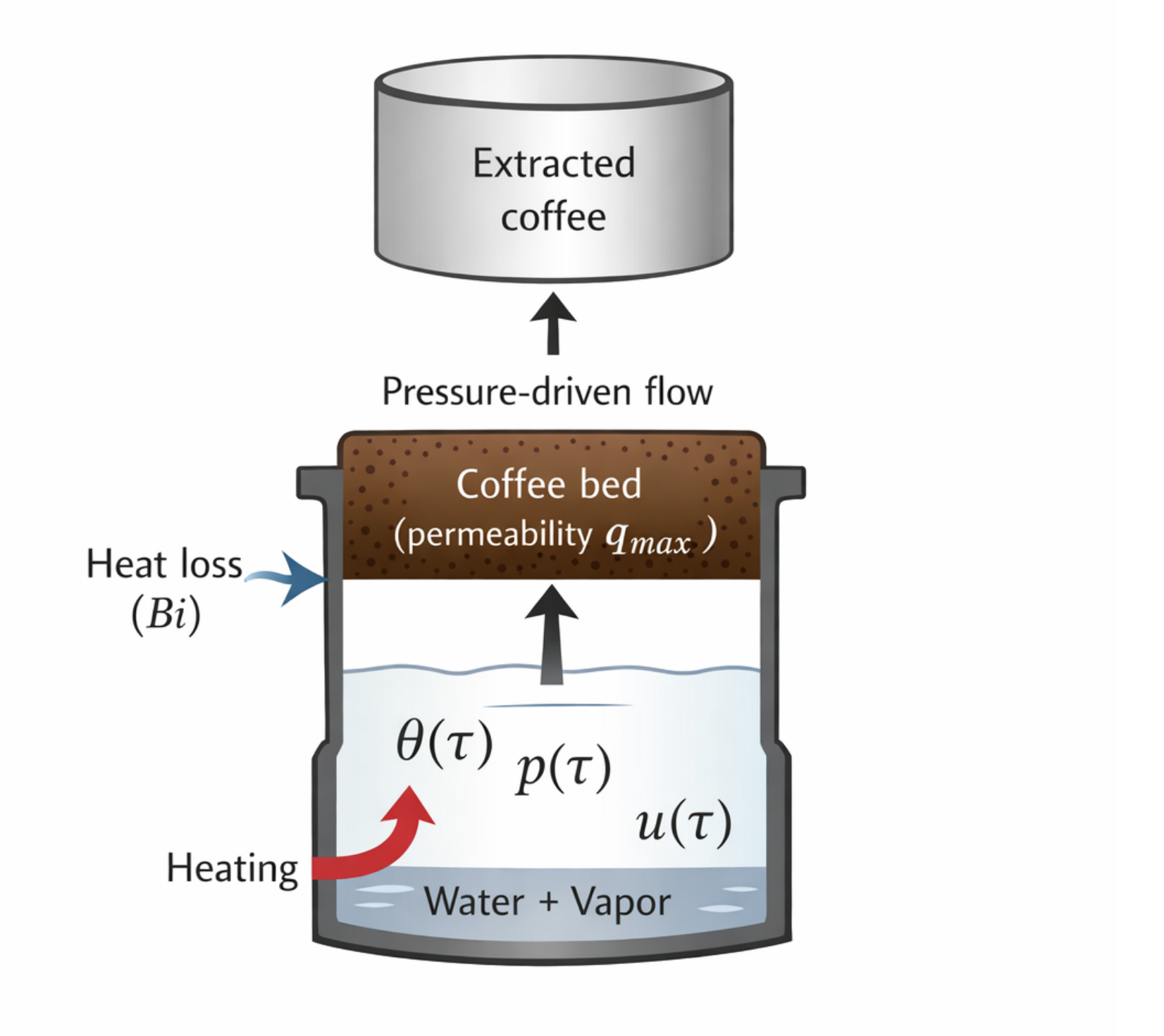}
    \caption{Schematic diagram of the moka pot and its reduced dynamical model. Heating of the lower chamber generates vapor pressure that increases the dimensionless temperature $\theta(\tau)$ and pressure $p(\tau)$. Once the pressure exceeds a threshold, liquid is driven upward through the porous coffee bed, characterized by a maximum permeability $q_{\mathrm{max}}$, producing a volumetric flow rate $q$ and extracted volume $u(\tau)$. Heat loss to the environment is represented by the Biot-like parameter $\mathrm{Bi}$.}
    \label{fig:schematic}
\end{figure}

In this work, we adopt a \emph{lumped-parameter, reduced-order modeling approach} in which the dominant physical processes are captured using a small set of coupled ordinary differential equations. This strategy is commonly employed in pedagogical and reduced-order thermo--fluid modeling~\cite{bossart2025science}. Rather than deriving a detailed conservation-law model with many state variables, we construct a minimal dimensionless system that retains the essential qualitative dynamics while remaining transparent and numerically tractable.

\subsection{State Variables}

The model is formulated in terms of three dimensionless, time-dependent state variables: the dimensionless temperature of the lower chamber, $\theta(\tau)$; the dimensionless pressure in the boiler, $p(\tau)$; and the cumulative dimensionless volume of extracted liquid, $u(\tau)$ (Fig.~\ref{fig:schematic}). The dimensionless time variable $\tau=t/t_0$ is scaled by a characteristic heating time $t_0$, corresponding to the timescale over which the boiler temperature rises appreciably in the absence of extraction. Temperature and pressure are scaled by characteristic values associated with the onset of pressure-driven flow, while $u$ is normalized by the initial liquid volume.

\subsection{Governing Equations}

The temporal evolution of the system is governed by the coupled dimensionless equations
\begin{align}
\frac{d\theta}{d\tau} &= 1 - \mathrm{Bi}\,\theta - \Lambda\,q, \label{eq:theta}\\
\frac{dp}{d\tau} &= \Lambda \left( e^{\theta} - p \right) - \Pi\,q, \label{eq:p}\\
\frac{du}{d\tau} &= q. \label{eq:u}
\end{align}

The parameter $\Lambda$ appears in both Eqs.~(\ref{eq:theta}) and (\ref{eq:p}), reflecting the shared thermodynamic origin of vapor generation and convective energy removal. Although denoted as $\mathrm{Bi}$, this parameter is not a classical Biot number but plays an analogous role by comparing heat loss to heating input in the lumped system. Extraction onset is defined as the time at which the dimensionless pressure first exceeds the threshold value $p=1$, corresponding to the onset of nonzero flow ($q>0$) through the coffee bed.

\subsubsection{Pre-extraction Heating Regime}

During the initial heating stage, the internal pressure remains below the extraction threshold and no liquid flow occurs ($q=0$). In this regime, Eq.~(\ref{eq:theta}) reduces exactly to
\begin{equation}
\frac{d\theta}{d\tau} = 1 - \mathrm{Bi}\,\theta,
\end{equation}
which governs the temperature evolution prior to extraction. This reduced equation is used to compare the model with experimental temperature--time data and to determine the characteristic heating timescale $t_0$. No pressure or flow parameters are fitted during this stage.

\subsubsection{Pressure Dynamics}

Equation~(\ref{eq:p}) describes the evolution of pressure in the lower chamber. The term $e^{\theta}$ represents the dimensionless equilibrium vapor pressure associated with the current temperature, capturing the leading-order exponential dependence predicted by the Clausius--Clapeyron relation near boiling~\cite{callen1993thermodynamics}. Pressure is assumed to relax dynamically toward this equilibrium value rather than being imposed instantaneously, allowing transient thermo--fluid feedback to be resolved without numerical stiffness.

\subsubsection{Extraction Flow}

The extraction flow rate is modeled using the thresholded linear relation
\begin{equation}
q = q_{\max}\,\max(p-1,0),
\end{equation}
representing pressure-driven flow through the porous coffee bed, consistent with Darcy-type models for flow in porous media~\cite{bear2013dynamics}. The threshold $p=1$ corresponds to the minimum pressure required to overcome hydrostatic head and flow resistance.

\subsection{Dimensionless Parameters}

The model contains four dimensionless parameters:
\begin{itemize}
    \item $\mathrm{Bi}$: heat-loss parameter,
    \item $\Lambda$: thermo--fluid coupling strength,
    \item $q_{\max}$: effective permeability of the coffee bed,
    \item $\Pi$: pressure dissipation coefficient.
\end{itemize}

These parameters define a physically interpretable parameter space within which distinct extraction behaviors emerge.

\subsection{Initial Conditions}

At the onset of heating,
\begin{equation}
\theta(0)=0, \quad p(0)=1, \quad u(0)=0.
\end{equation}

\begin{table}[h]
\centering
\caption{Numerical values of dimensionless parameters used in the simulations.}
\begin{tabular}{lll}
\hline
\textbf{Parameter} & \textbf{Value} & \textbf{Description} \\
\hline
$\mathrm{Bi}$ & $0.15$ 
& Heat loss coefficient \\[4pt]

$\Lambda$ & $1.5$ 
& Vapor--pressure coupling strength \\[4pt]

$\Pi$ & $0.5$ 
& Pressure depletion due to extraction \\[4pt]

$q_{\max}$ & $1.2$ 
& Maximum dimensionless flow rate \\[4pt]

\hline
\end{tabular}
\label{tab:regimes}
\end{table}

\subsection{Numerical Implementation}

The coupled equations were solved using a variable-step explicit Runge--Kutta method (\texttt{solve\_ivp}, SciPy). Relative and absolute tolerances of $10^{-6}$ and $10^{-8}$ were used. Parameter values employed in representative simulations are listed in Table~\ref{tab:regimes}.

\subsection{Model Limitations}

The model is intentionally simplified. Spatial temperature gradients, two-phase flow effects, and time-dependent permeability changes are neglected. The vapor pressure is represented using a minimal exponential form, and the flow law is linearized for pedagogical clarity. Consequently, the model is not intended for device-specific optimization but rather for illustrating how coupled thermo--fluid dynamics can be modeled, simplified, and tested against experimental data.

\section{Results}

\begin{figure}
    \centering
    \includegraphics[width=1\linewidth]{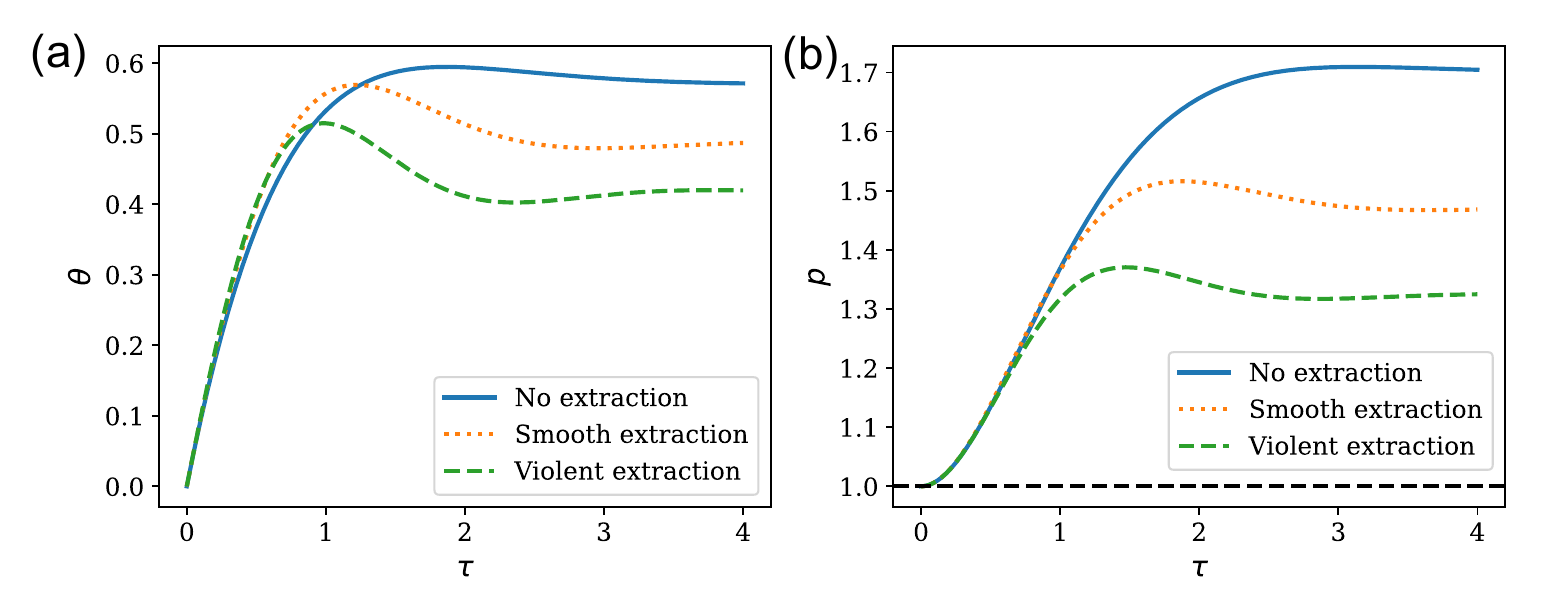}
    \caption{Dimensionless temperature $\theta(\tau)$ (left) and pressure $p(\tau)$ (right) for three representative extraction regimes. In the absence of extraction, thermal energy accumulates in the boiler, leading to higher steady-state temperature and pressure. When pressure-driven flow occurs, extraction removes both mass and energy, reducing peak temperature and pressure. The dashed line indicates the pressure threshold for flow onset. The three curves correspond to the parameter values listed in Table~\ref{tab:regimes}.}

    \label{fig:regimes}
\end{figure}

\subsection{Dimensionless Temperature and Pressure Dynamics}

Figure~\ref{fig:regimes} shows the dimensionless temperature $\theta(\tau)$ and pressure $p(\tau)$ for three representative extraction regimes obtained by varying the Biot-like heat-loss parameter $\mathrm{Bi}$ and the effective permeability parameter $q_{\max}$, while all other parameters are held fixed. In all cases, the temperature initially increases due to external heating, leading to vapor generation and a gradual rise in pressure within the lower chamber.

In the absence of significant extraction (no-extraction regime), pressure remains close to the threshold value and thermal energy accumulates in the boiler, resulting in the highest steady-state temperature. When pressure-driven flow occurs, extraction removes both mass and energy from the lower chamber, limiting further increases in temperature and pressure. This effect is most pronounced in the rapid extraction regime, where strong thermo--fluid coupling produces an early pressure peak followed by partial relaxation.

These results demonstrate the strong coupling between thermal and fluid processes in the moka pot and illustrate how extraction acts as a self-regulating mechanism that limits temperature and pressure growth once flow is initiated.

\subsection{Flow Rate Dynamics and Extraction Completion}

\begin{figure}
    \centering
    \includegraphics[width=1\linewidth]{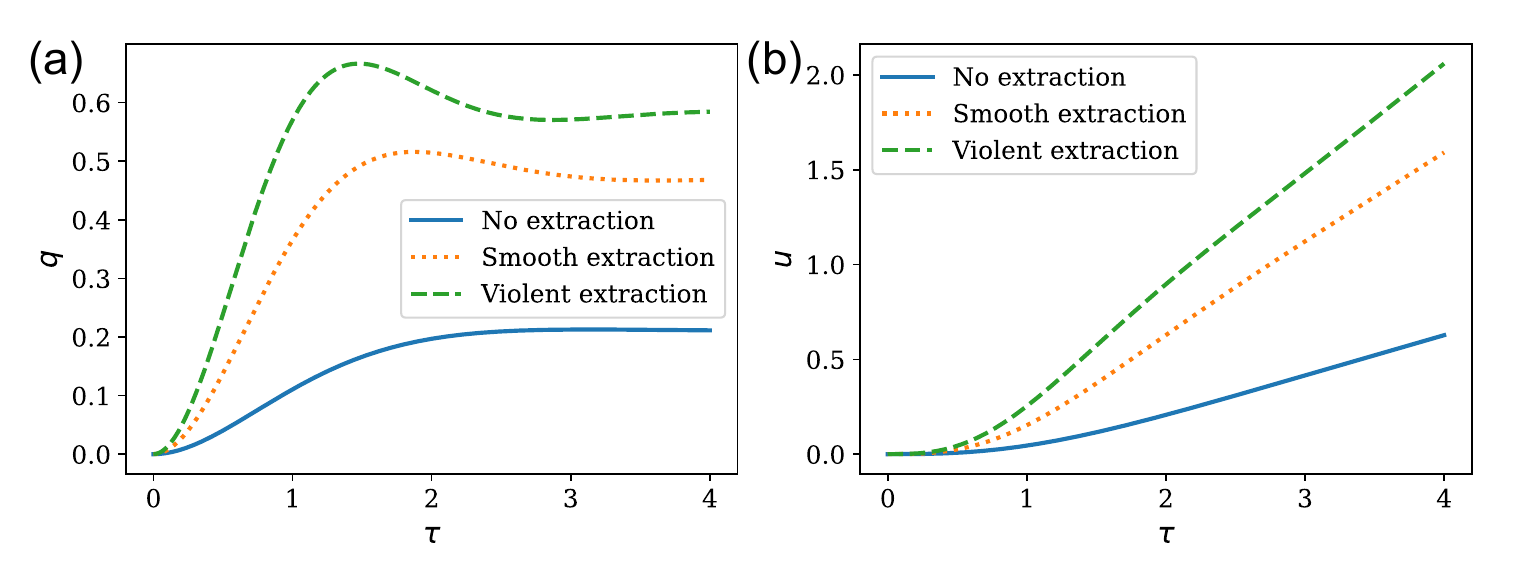}
    \caption{Dimensionless extraction flow rate $q(\tau)$ (left) and cumulative extracted volume fraction $u(\tau)$ (right). Flow begins only after the pressure exceeds a threshold value and exhibits a rapid rise followed by relaxation. The extracted volume displays a characteristic S-shaped evolution, marking the transition from heating-dominated to flow-dominated dynamics.}

    \label{fig:flow_volume}
\end{figure}

The corresponding extraction flow rates $q(\tau)$ and cumulative extracted volume fractions $u(\tau)$ are shown in Fig.~\ref{fig:flow_volume}. Flow begins only after the pressure exceeds the threshold value $p=1$, reflecting the need to overcome hydrostatic head and resistance in the coffee bed. Following onset, the flow rate increases rapidly before reaching a maximum and subsequently relaxing toward a quasi-steady value.

In the rapid extraction regime, the flow rate exhibits an early peak due to rapid pressure buildup, followed by a decline as pressure relaxes and thermal driving diminishes. In contrast, the smooth extraction regime displays a more gradual increase and sustained flow, resulting in a longer extraction time. The no-extraction regime shows only weak flow and incomplete extraction over the simulated interval.

The cumulative extracted volume displays a characteristic S-shaped evolution, marking the transition from heating-dominated to flow-dominated dynamics. This behavior emerges naturally from the coupled thermo--fluid model and is commonly observed during household moka pot brewing.

\begin{figure}
    \centering
    \includegraphics[width=1\linewidth]{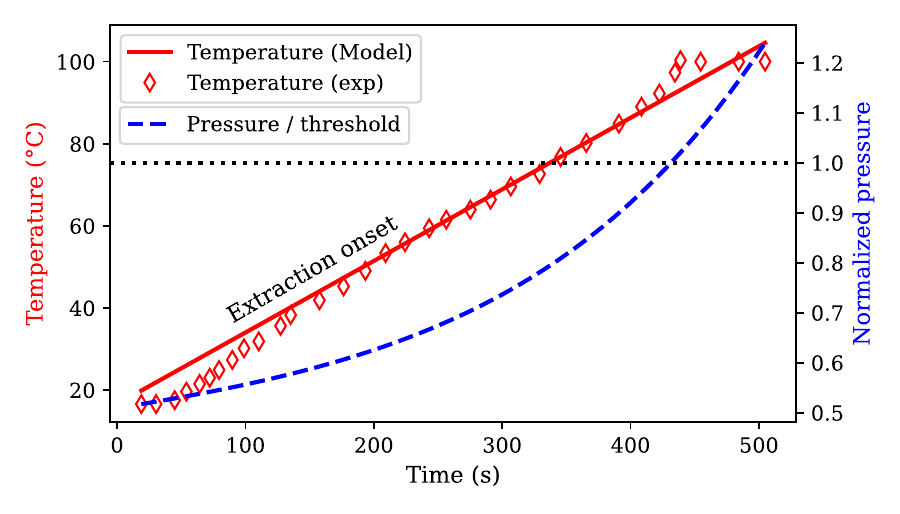}
\caption{Experimental temperature data from Ref.~\cite{gianino2007experimental} (symbols) and model prediction (solid line) during the heating stage of a moka pot. The normalized pressure (blue dashed line, right axis) is computed from the experimentally constrained temperature. The extraction threshold ($p=1$) defines the predicted onset of extraction.}
    \label{fig:experiment}
\end{figure}

\subsection{Comparison with Experimental Temperature and Predicted Pressure}

To assess the physical relevance of the minimal model, its heating-stage predictions are compared with published experimental temperature--time data for a moka pot. During the initial heating stage, the internal pressure remains below the extraction threshold and no flow occurs ($q=0$), reducing the temperature equation to a single ordinary differential equation. This reduced equation is fitted to experimental temperature data to determine the characteristic heating timescale $t_0$.

Figure~\ref{fig:experiment} shows the experimentally measured temperature evolution together with the model prediction. The agreement is good throughout the heating stage up to the onset of boiling. Using the experimentally constrained temperature evolution as input, the model is then used to compute the pressure rise inside the boiler by combining contributions from compressed air and water vapor.

The resulting pressure evolution is shown on a separate axis in Fig.~\ref{fig:experiment}. The pressure remains below the extraction threshold during early heating and rises rapidly near boiling, crossing the threshold at $t \approx 4.3 \times 10^2~\mathrm{s}$. This crossing defines the predicted onset of extraction. Importantly, the pressure evolution and extraction onset are not independently fitted but follow directly from the experimentally constrained temperature.

\section{Discussion}

\subsection{Physical Interpretation of the Extraction Dynamics}

The minimal model developed in this work highlights the central role of nonlinear feedback between heating, vapor pressure generation, and pressure-driven flow in governing moka pot extraction. As the lower chamber is heated, the temperature rises and induces an exponential increase in vapor pressure. Once the pressure exceeds a threshold set by hydrostatic head and flow resistance, liquid is forced through the coffee bed, initiating extraction.

The onset of extraction fundamentally alters the system dynamics. Pressure-driven flow removes both mass and thermal energy from the boiler, introducing a negative feedback that limits further increases in temperature and pressure. This feedback mechanism explains the self-regulating character of moka pot operation and the finite duration of extraction observed in practice.

Within this framework, the model naturally predicts three qualitative extraction regimes. When permeability is low or heat loss is strong, pressure remains near the threshold and extraction is delayed or negligible. For intermediate parameters, a smooth extraction regime emerges, characterized by sustained pressure-driven flow and gradual volume increase. When thermo--fluid coupling is strong and flow resistance is low, rapid extraction occurs, marked by an early pressure peak followed by relaxation as extraction depletes the driving pressure. The term ``rapid'' (or ``violent'') extraction is used descriptively to denote strongly coupled pressure--flow dynamics; no mechanical instability is implied.

These regimes arise from the intrinsic nonlinear structure of the coupled thermo--fluid system rather than from detailed geometric or material properties of a specific device.

\subsection{Relation to Prior Work and Experimental Observations}

Although deliberately simplified and nondimensional, the present model is consistent with prior analyses and experimental observations of moka pot operation. In particular, the delayed onset of flow, the existence of a finite extraction time, and the sensitivity of extraction dynamics to heating and flow resistance agree qualitatively with experimental studies and with the thermofluid analysis developed by King~\cite{king2008physics}.

Compared to dimensional balance-law models, the present approach emphasizes a reduced, fully dimensionless description that highlights dynamical regimes and feedback mechanisms. The two approaches are therefore complementary: detailed models provide physical grounding and quantitative interpretation, while the reduced framework presented here offers conceptual clarity and pedagogical accessibility.

The comparison with experimental temperature data further clarifies this distinction. The heating dynamics are constrained directly by measurement, while the pressure evolution and extraction onset emerge as genuine model predictions. This separation between calibration and prediction illustrates how simplified models can be tested against limited experimental data without sacrificing interpretability.

\subsection{Pedagogical Implications and Scope}

Beyond its relevance to coffee preparation, the moka pot provides a compact pedagogical example of coupled nonlinear thermo--fluid dynamics in an everyday physical system. Because the model consists of only three coupled ordinary differential equations with physically interpretable parameters, it is well suited for classroom use, computational laboratories, and student-led exploration.

The dimensionless formulation enables systematic parameter studies that illustrate threshold phenomena, feedback loops, relaxation dynamics, and self-regulation. The model can be readily implemented using standard numerical tools and extended incrementally to explore additional physical effects, making it particularly suitable for advanced undergraduate instruction.

In this sense, the moka pot serves as a familiar and engaging platform through which abstract concepts in thermodynamics, fluid mechanics, and nonlinear dynamics can be connected to everyday experience.

\section{Conclusion}

We have presented a minimal, fully dimensionless thermo--fluid model of moka pot operation that captures the essential coupling between heating, vapor pressure generation, and pressure-driven extraction. By reducing the system to a small set of ordinary differential equations with physically interpretable parameters, the model provides a transparent framework for understanding the nonlinear feedback mechanisms governing extraction dynamics.

A key result of this work is the explicit confrontation of the model with experimental data. The heating-stage temperature evolution is quantitatively compared with published measurements, allowing the characteristic thermal timescale to be fixed independently. Using the experimentally constrained temperature as input, the model then predicts the pressure rise and the onset of extraction without additional fitting parameters. The predicted extraction onset time is consistent with experimentally observed heating timescales, demonstrating that the minimal model captures the dominant physics governing moka pot operation.

Despite its simplicity, the model reproduces several qualitatively distinct extraction regimes, including delayed or negligible extraction, smooth extraction, and rapid extraction driven by strong thermo--fluid coupling. These regimes arise naturally from the nonlinear structure of the governing equations rather than from device-specific tuning. While the model does not attempt detailed quantitative prediction for specific moka pot designs, it provides physically meaningful trends and timescales that are relevant for both experimental comparison and conceptual understanding.

Beyond coffee preparation, the moka pot serves here as a compact pedagogical example of how physicists construct, simplify, and test models of coupled nonlinear systems using limited experimental input. The framework is readily implementable with standard numerical tools and is well suited for instructional use in thermodynamics, fluid mechanics, and computational physics courses. More detailed models incorporating spatial variation, two-phase flow, or refined constitutive relations may be developed within this framework, but the present work demonstrates that even a highly reduced description can yield significant physical insight.

%\begin{acknowledgments}
%This research was supported by the Directorate of Research and Development, Universitas Indonesia, under the Hibah Riset FMIPA UI 2023 (Grant No. PKS-028/UN2.F3.D/PPM.00.02/2023). The author thanks Prof. Terry Mart and Dr. Arief Sudarmadji of Universitas Indonesia for fruitful discussions, as well as for supporting the department with coffee and milk.
%end{acknowledgments}

\end{document}